\documentclass[twocolumn,showpacs,preprintnumbers]{revtex4}
\usepackage{amssymb}
\usepackage{amsmath}
\usepackage{graphicx}
\usepackage{dcolumn}
\usepackage{bm}
\usepackage{subfigure}

\begin{document}

\title{Perfect photon absorption in hybrid atom-optomechanical system}
\author{Yang Zhang}
\author{Chang-shui Yu}
\email{quaninformation@sina.com;ycs@dlut.edu.cn}
\affiliation{School of Physics and Optoelectronic Technology, Dalian University of
Technology, Dalian 116024, China }

\begin{abstract}


Recently, the photon absorption attracts lots of interest and  plays an important role in a variety of applications. Here, we propose a valuable scheme to investigate the perfect photon absorption
in a hybrid atom-optomechanical system both under and beyond the low-excitation limit.
 The perfect photon absorption persists both in the linear atomic excitation
regime and nonlinear atomic excitation regime,  below the threshold of the optical bistability/multistability, respectively.  We also show that the optical nonlinearity raised by the
nonlinear optomechanical interaction and nonlinear atomic excitation can be overlap-added, there presents
a perfect corresponding relation between perfect photon absorption and the optical multistability beyond the low-excitation limit, the optical bistability can be switched to the optical multistability by increasing the input intensity.
The combination of the perfect photon absorption and optical
bistability/mutistability is useful for the photon switch application. We believe that
this study will provide a possible design of an optical switch.
\end{abstract}

\maketitle


\section{Introduction}

It is well known that photons as the carrier of the information, play an
important role in quantum information and quantum communication.
Recently, the photon absorption attracts a great deal of attention and has aroused
widespread interest in recent studies \cite{wan,stone,MK,huang}. Normally,
the coherent perfect absorption corresponding to a certain frequency 
is determined by the intrinsic properties of the medium and the absorption of the input light was reported in Ref. \cite{cupta1,cupta}. Recently, the coherent perfect photon absorption using path entanglement was also
theoretically studied in Ref. \cite{huang} and then was demonstrated experimentally
by Faccio et al \cite{Faccio}. The physical basis behind it is the destructive interference between the two input fields \cite%
{stefano,cupta}. It is shown that the perfect photon absorption has potential applications in optics communications and  photonic devices including transducers, modulators, and optical switches and transistors \cite{Agarwal-arxiv, stone,12-19}.
Therefore,  some fundamental efforts have also been made on the achievement and applications of perfect photon absorption in various systems such as  whispering-gallery-mode micro-resonators \cite{chow}, the
coupled atom-resonator-waveguide system \cite{fan}, second harmonic
generation \cite{chen}, nanostructured graphene film \cite{jinfa} and strongly scattering media \cite{stone2} and so on.
In particular, the perfect photon absorption has also been studied not only in the cavity quantum electrodynamics \cite{Agarwal-PRA,Agarwal-arxiv} but also in the cavity optomechanical system (COM) which enables the coupling between the mechanical modes and the
optical field via the radiation pressure and attracts a lot of
interest due to the rich physics \cite{vaha,modern} and applications, e.g., ground state
cooling \cite{cooling}, quantum coherent state transfer \cite{ Palo},
ponderomotive squeezing \cite{safa}, quantum entanglement \cite{vita},
optical Kerr nonlinearity \cite{rabl}, and gravitational wave physics  \cite{ligo,abbott}. However, is the optomechanical coupling necessary
in the perfect photon absorption in COM? or what's the essential role of the optomechancial coupling in the perfect photon absorption?


In this paper, we study the perfect photon absorption as well we the optical bistability/multistability in a hybrid atom-optomechanical
system which couples to an ensemble of two-level atoms. Our
results show that,  the perfect photon absorption can be implemented in this system but always accompanied with the bistability or multistability of the output fields. while the optical
intensity onset the bistability. In difference with Re \cite{Agarwal-arxiv},
where the optical nonlinearity relies on the nonlinear atomic excitation,
here the optical nonlinearity used is arisen from the radiation pressure
interaction between moveable mirror and the cavity field, meanwhile the
atomic ensemble linearily coupled with the cavity under the low excitation
limit. Moreover, the photon absorption and optical bistability can be
modified by the interference between the two input fields. In addition, we also
employ both the atomic nonlinear excitation and nonlinear optomechanical
interaction, the two nonlinearity can be overlap-add, both the intra-field
and output field intensity can exhibit the optical multistability, the optical
bistability switches to the optical multistability by increasing the input
field intensity. Furthermore, the perfect absorption point also can be
increased with the help of the optomechanical interaction. Finally, the
demonstration of realizable within the current experimental technology on
our scheme is also been discussed.






\section{The model}

\begin{figure}[tbp]
\centering
\includegraphics[width=0.8\columnwidth,height=1.5in]{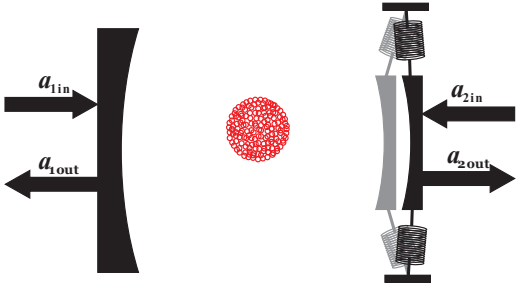}
\caption{(color online) Schematic setup. The hybrid atom-optomechanical
system consists of a cloud of two level atoms in the cavity which is driven
by two input light fields.}
\end{figure}
\begin{figure}[tbp]
\centering
\hspace*{-2.2cm} \includegraphics[width=1.5%
\columnwidth,height=2.5in]{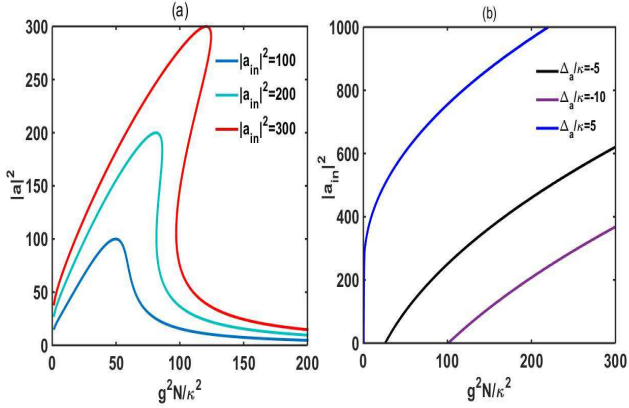}
\caption{(color online) Under the perfect absorption condition Eq. (\protect
\ref{perfect11}) and Eq. (\protect\ref{perfect12}), (a) intra-cavity light
intensity $\left\vert a\right\vert ^{2}$ vs the fitting parameter $g^{2}N$
with the different input light intensity $\left\vert a_{in}\right\vert ^{2}$%
, the cavity-laser detuning $\Delta _{a}/\protect\kappa =-5$. (b) the input
intensity as a function of the fitting parameter $g^{2}N$ with the different
cavity-laser detuning. The other system parameters are $\Gamma /\protect%
\kappa =2$, $\protect\omega _{m}/\protect\kappa =0.01$, $\protect\gamma /%
\protect\kappa =0.1$, $g_{0}/\protect\kappa =0.1$, $\protect\delta /\protect%
\kappa =-\protect\sqrt{g^{2}N/\protect\kappa ^{2}-1}$.}
\end{figure}
As schematically shown in Fig. 1, we consider a typical hybrid
atom-optomechanical system which consists of a bare optical cavity coupled
to a mechanical mode and $N$ identical two-level atoms trapped in the
cavity. The atomic transition frequency and the atomic line width are
denoted by $\omega _{e}$ and $\Gamma $, respectively. In the rotating frame
of the input laser frequency, the full Hamiltonian with regard for the
driving and the dissipation, is given by (set $\hbar =1$ hereafter)
\begin{eqnarray}
H &=&\Delta _{a}a^{\dag }a+\delta S^{z}+\omega _{m}b^{\dag
}b+g(aS^{+}+a^{\dag }S^{-})  \label{hamiltonian} \\
&&-g_{0}a^{\dag }a(b^{\dag }+b)+i\sqrt{2\kappa _{1}}a_{1in}a^{\dag }+i\sqrt{%
2\kappa _{2}}a_{2in}a^{\dag }  \notag \\
&&-i\sqrt{2\kappa _{1}}a_{1in}^{\dag }a-i\sqrt{2\kappa _{2}}a_{2in}^{\dag }a,
\notag
\end{eqnarray}%
where $a$ denotes the cavity mode, and $S^{z}=\frac{1}{2}\sum_{i=1}^{N}(%
\sigma _{i}^{+}\sigma _{i}^{-}-\sigma _{i}^{-}\sigma _{i}^{+})$ and $S^{\pm
}=\sum_{i=1}^{N}\sigma _{i}^{\pm }$ stand for the collective atomic
operators with $\sigma _{i}^{\pm }=\left\vert e\right\rangle
_{i}\left\langle g\right\vert $ representing the $i$th atomic raising and
lowering operators. Here $a_{1in}$ and $a_{2in}$ are two input fields upon
the cavity with the same frequency $\omega _{l}$. $\Delta _{a}=\omega
_{a}-\omega _{l}$ is the laser detuning from the cavity mode and $\delta
=\omega _{e}-\omega _{l}$ is the laser detuning from the atoms. Based on
Hamiltonian (\ref{hamiltonian}) and the potential dissipation processes, one
can derive the dynamics of the system from the Heisenberg equations of the
operators. In the classical limit, namely, we drop the quantum fluctuations
and replace the operators by their expectation values, one can obtain the
equations of motion of the expectation values of the operators as follows.
\begin{equation}
\left\langle \dot{b}\right\rangle =-i\omega _{m}\left\langle b\right\rangle
-\gamma \left\langle b\right\rangle -ig_{0}\left\langle a^{\dag
}a\right\rangle ,  \label{lan1}
\end{equation}%
\begin{equation}
\left\langle \dot{S}^{-}\right\rangle =-i\delta \left\langle
S^{-}\right\rangle +2ig\left\langle aS^{z}\right\rangle -\frac{\Gamma }{2}%
\left\langle S^{-}\right\rangle ,
\end{equation}%
\begin{equation}
\left\langle \dot{S}^{z}\right\rangle =-\Gamma \left\langle
S^{z}\right\rangle -ig\left\langle aS^{+}\right\rangle +ig\left\langle
a^{\dag }S^{-}\right\rangle -\frac{N\Gamma }{2},
\end{equation}%
\begin{eqnarray}
\left\langle \dot{a}\right\rangle &=&-i\Delta _{a}\left\langle
a\right\rangle -\kappa \left\langle a\right\rangle -ig\left\langle
S^{-}\right\rangle -ig_{0}\left\langle a\right\rangle (\left\langle b^{\dag
}\right\rangle  \notag \\
&&+\left\langle b\right\rangle )+\sqrt{2\kappa _{1}}a_{1in}+\sqrt{2\kappa
_{2}}a_{2in},  \label{lan2}
\end{eqnarray}%
where $\left\langle \cdots \right\rangle $ represents the expectation value
over the steady state. Combining above equations, we can solve the
steady-state solution and obtain the intra-cavity field as
\begin{equation}
\left\langle a\right\rangle =\frac{\sqrt{2\kappa _{1}}a_{1in}+\sqrt{2\kappa
_{2}}a_{2in}}{i\Delta _{a}+\kappa -\frac{g^{2}N}{\frac{N}{2\left\langle
S^{z}\right\rangle }(\frac{\Gamma }{2}+i\delta )}-i\Xi \left\vert
a\right\vert ^{2}},  \label{non}
\end{equation}%
where $\left\langle S^{z}\right\rangle =-\frac{N}{2(1+\frac{2g^{2}\left\vert
a\right\vert ^{2}}{\frac{\Gamma ^{2}}{4}+\delta ^{2}})}$ and $\Xi =\frac{2\omega
_{m}g_{0}^{2}}{\gamma ^{2}+\omega _{m}^{2}}$. From Eq. (\ref{non}), one can
easily get the intra-cavity intensity $\left\vert a\right\vert ^{2}$ and,
hence, obtain the properties of the photon absorption.

\section{The photon absorption}

\subsection{ Perfect photon absorption in the low-excitation limit}

We first consider that the atomic ensemble is only in the low excitation
regime. In this case, one can set $\left\langle S^{z}\right\rangle \approx -%
\frac{N}{2}$, so Eq. (\ref{non}) can be reduced to
\begin{equation}
\left\langle a\right\rangle =\frac{\sqrt{2\kappa _{1}}a_{1in}+\sqrt{2\kappa
_{2}}a_{2in}}{i\Delta _{a}+\kappa +\frac{g^{2}N}{\frac{\Gamma }{2}+i\delta }%
-i\Xi \left\vert a\right\vert ^{2}}.  \label{aalinear}
\end{equation}%
It is obvious that Eq. (\ref{aalinear}) includes the intra-cavity intensity $%
\left\vert a\right\vert ^{2}$ and demonstrates the nonlinear dependence of
the intra-cavity intensity. So the system may exhibit the optical
bistability under the certain parameter range. In order to find the steady
states of the two output light fields, one will have to consider the
following input-output relation \cite{wall}
\begin{equation}
a_{1out}=\sqrt{2\kappa _{1}}a-a_{1in},  \label{inout1}
\end{equation}%
\begin{equation}
a_{2out}=\sqrt{2\kappa _{2}}a-a_{2in}.  \label{inout2}
\end{equation}%
One can assume that $a_{1in}=\left\vert a_{in}\right\vert $, $%
a_{2in}=e^{i\varphi }\left\vert a_{in}\right\vert $, where $\varphi $
denotes the relative phase of the two opposite input fields. We also consider
a symmetric cavity with $\kappa _{1}=\kappa _{2}=\frac{\kappa }{2}$, then
the output intensity can be calculated as
\begin{eqnarray}
\frac{\left\vert a_{1out}\right\vert ^{2}}{\left\vert a_{in}\right\vert ^{2}}
&=&\left\vert \frac{\kappa (1+e^{i\varphi })}{i\Delta _{a}+\kappa +\frac{%
g^{2}N}{\frac{\Gamma }{2}+i\delta }-i\Xi \left\vert a\right\vert ^{2}}%
-1\right\vert ^{2},  \label{linear1} \\
\frac{\left\vert a_{2out}\right\vert ^{2}}{\left\vert a_{in}\right\vert ^{2}}
&=&\left\vert \frac{\kappa (1+e^{i\varphi })}{i\Delta _{a}+\kappa +\frac{%
g^{2}N}{\frac{\Gamma }{2}+i\delta }-i\Xi \left\vert a\right\vert ^{2}}%
-e^{i\varphi }\right\vert ^{2}.  \label{linear2}
\end{eqnarray}

Let us now focus our attention on how to realize the perfect photon
absorption. The perfect photon absorption implies $a_{1out}=a_{2out}=0$. One
can find that if $\varphi \neq 2k\pi $, $k=0,\pm 1,\pm 2,\cdots $, the
perfect photon absorption will not happen at any rate. So we restrict
ourselves under the condition $\varphi =0$ without loss of generality. Thus
we can get two specific conditions for the perfect absorption:
\begin{equation}
\frac{\kappa }{\Gamma }=\frac{2g^{2}N}{\Gamma ^{2}+4\delta ^{2}},
\label{perfect11}
\end{equation}%
\begin{equation}
(\frac{\kappa \Gamma }{2}+\Delta _{a}\delta -g^{2}N)=\delta \Xi \left\vert
a\right\vert ^{2}.  \label{perfect12}
\end{equation}%
In terms of Eqs. (\ref{perfect11}) and (\ref{perfect12}), we discuss some
pertinent results about the perfect absorption and the bistable behaviors in
the linear atomic excitation regime. By comparing with Ref. \cite{Agarwal-PRA},
one can find that the two perfect trapping conditions presented in Ref. \cite%
{Agarwal-PRA} can be summarized by our Eq. (\ref{perfect11}). However, Eq. (%
\ref{perfect12}) is our distinguishing condition which is the result of the
optomechanical interaction. The nonlinear interaction of the
cavity-mechanical plays the dominant role in the optical bistability, and
the hybrid atom-optomechanical system is driven into the bistable domain
when the input light intensity is above the threshold for the onset of the
optical bistability \cite{jiangcheng}. As shown in Eq. (\ref{aalinear}), the
intra-cavity photon number is determined by both the atomic ensemble and the
optomechanical parameters, thus we can harness them to control the optical
bistable behavior and the photon absorption. For example, we can effectively
control the parameters $g^{2}N$ and the atomic frequency to match Eq. (\ref%
{perfect11}) and control the intra-cavity intensity to match Eq. (\ref%
{perfect12}).
\begin{figure}[tbp]
\centering
\hspace*{-0.5cm} \subfigure{\includegraphics[width=0.55%
\columnwidth,height=2in]{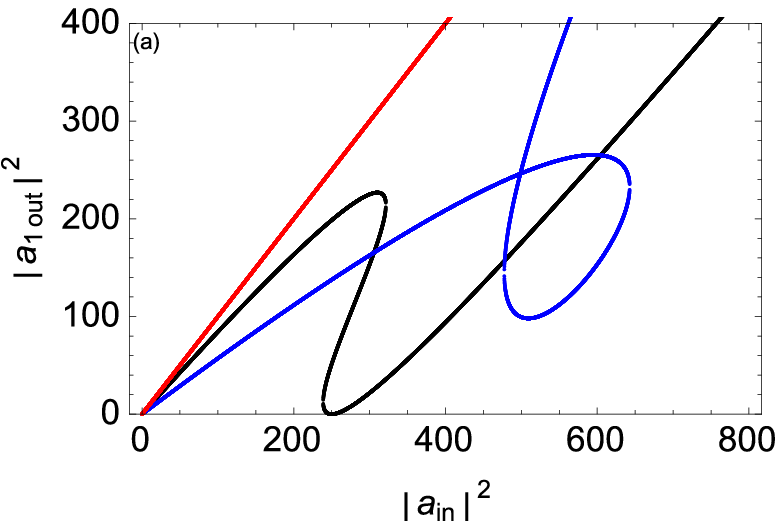}}\subfigure{\includegraphics[width=0.55%
\columnwidth,height=2in]{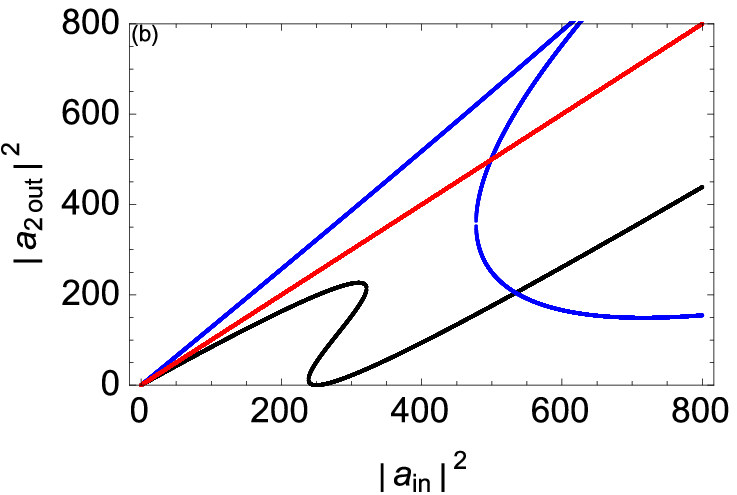}}\newline
\caption{(color online).Output light intensity $\left\vert
a_{out}\right\vert ^{2}$ as a function of input light intensity $\left\vert
a_{in}\right\vert ^{2}$ with the different relative phase $\protect\varphi $%
. For this case, the two input lights have the same intensity $\left\vert
a_{1in}\right\vert ^{2}$ =$\left\vert a_{2in}\right\vert ^{2}$=$%
\left\vert a_{in}\right\vert ^{2}$. The black line is for $\protect\varphi %
=0 $, the blue curve stands for $\protect\varphi =\protect\pi /2$, the red
line stands for $\protect\varphi =\protect\pi $. The other parameters are
the same with Fig. 2. }
\end{figure}
In Fig. 2 (a) and Fig. 2 (b) we display the intra-cavity intensity $%
\left\vert a\right\vert ^{2}$ and the input light intensity $\left\vert
a_{in}\right\vert ^{2}$ as a function of the fitting parameter $%
g^{2}N/\kappa ^{2}$ under the condition of the perfect photon absorption.
From Fig. 2 (a), one can see that the $\left\vert a\right\vert ^{2}$
exhibits the bistability if the input intensity is large enough. In
addition, even though the large input intensity can make the optomechanical
system show the apparent bistability, this could require a relatively strong
coupling $g^{2}N/\kappa ^{2}$, which is also illustrated by the monotonic
relation between $\left\vert a_{in}\right\vert ^{2}$ and $g^{2}N/\kappa ^{2}$
in Fig. 2 (b). It is also shown in Fig. 2 that the requirement of strong
coupling can actually be compensated for by adding more atoms, or weakening
the leakage $\kappa $ or reducing the detuning $\Delta _{a}/\kappa $. In
addition, for nontrivial solutions, $\left\vert a_{in}\right\vert >0$
requires that besides the condition given in Eq. (\ref{perfect11}), the
perfect absorption should only occur in the following range of $g^{2}N$,
that is, (1) $g^{2}N>\frac{\Delta _{a}^{2}+\kappa \Gamma +\sqrt{\Delta
_{a}^{4}+2\kappa \Gamma \Delta _{a}^{2}-4\Delta _{a}^{2}}}{2}$ for $\Delta
_{a}<0$; (2) $g^{2}N>\frac{\kappa \Gamma }{2}$ for $\Delta _{a}>0$. The
numerical illustration of these ranges is given in Fig. 2 (b). Both the
ranges imply the strong coupling $g^{2}N$. In addition, one can also find
that the large atomic frequency detuning $\Delta _{a}<0$ will lead to the
stronger coupling.

In what follows, we present the result for the two output light intensities $%
\left\vert a_{1out}\right\vert ^{2}$ and $\left\vert a_{2out}\right\vert
^{2} $ for the hybrid system in Fig. 3 (a) and Fig. 3 (b). One can find
that the perfect photon absorption occurs solely at a particular input
intensity $\left\vert a_{in}\right\vert ^{2}=\kappa \delta \Xi /(\frac{%
\kappa \Gamma }{2}+\Delta _{a}\delta -g^{2}N)$ with the relative phase $%
\varphi =2n\pi $, $n=0,\pm 1,\pm 2\cdots $ (the black lines in the Fig. 3).
Interestingly, we find that once the perfect absorption occurs, the output
field intensity behaves the optical bistability where we can get three real
distinct values for the output field intensity $\left\vert
a_{out}\right\vert ^{2}$ due to the nonlinear equation Eq. (\ref{aalinear}).
Based on the Eq. (\ref{perfect11}) and Eq. (\ref{perfect12}), one can find
that the perfect absorption can't occur if $g=0$; Similarly, the optical
bistability will vanish if $g_{0}=0$, even though the perfect absorption
could still be present, which is consistent with Ref. \cite{Agarwal-PRA}. In this
sense, the perfect photon absorption mainly relies on the coupling between
the atomic ensemble and the cavity, whereas the nonlinear optomechanical
coupling plays the dominant role in the optical bistability \cite{jiangcheng}. 
It is different from Ref. \cite{Agarwal-arxiv} where the optical
bistability relies on the nonlinear atomic excitation. In addition, the two
output intensities can also be modified by the interference of the two
identical input lights, which can be shown by Fig. 3 (a) and (b). In Fig. 3,
the blue lines correspond to $\varphi =\pi /2$ and the red lines correspond
to $\varphi =\pi $. One can find that the bistability disappears ($\varphi
=\pi $) or the system is driven to the complicated bistable domain ($\varphi
=\pi /2$) by the effect of the relative phase, but the perfect absorption is
absent. In other words, with $\varphi $ changing from $\pi $ to $0$, the
hybrid optomechanical system is driven from the monostable domain to the
complicated bistable domain. During this procedure, one particular
bistability results in the perfect absorption as shown in Fig. 3.

In Fig. 4, we plot the output light intensities versus $\left\vert
a_{in}\right\vert^2$ for the different cavity-laser detunings. One can see
that the different values of the atom-cavity detuning ($\Delta =\Delta
_{a}-\delta $) don't take the edge off the perfect photon absorption and the
optical bistability. The different $\Delta $ requires the different input
light intensities to match the special conditions of the perfect absorption.
With the small $\Delta $, the system can be easily driven to the nonlinear
regime and achieve the optical bistability, meanwhile, the perfect photon
absorption can also happen with the relatively weak input light intensity.

Fig. 3 (a) and Fig. 3 (b) have shown that the relative phase has a deep
influence on the photon absorption and the optical bistability. As mentioned
above, the hybrid atom-optomechanical system could be out of the bistable
domain and in particular, the perfect photon absorption does not occur due
to the non-vanishing effect of the relative phase. To give an intuitive
illustration, it is imperative to plot the output light intensity versus the
relative phase $\varphi $. In Fig. 5 we plot the two output light
intensities $\left\vert a_{1out}\right\vert ^{2}$ and $\left\vert
a_{2out}\right\vert ^{2}$ with $\left\vert a_{1in}\right\vert
^{2}=\left\vert a_{in}\right\vert ^{2}$, $\left\vert a_{2in}\right\vert
^{2}=e^{i\varphi }\left\vert a_{in}\right\vert ^{2}$. One can see that the
two output fields are not equal to each other except at some particular $%
\varphi $. These $\varphi $ can be determined by solving Eq. (\ref{linear1})
and Eq. (\ref{linear2}) associated with Eq. (\ref{aalinear}), and thus one
can have $\sin {\varphi }=0$ or $\cos \varphi =c=(\kappa +\frac{g^{2}N\Gamma
}{2(\frac{\Gamma ^{2}}{4}+\delta ^{2})})^{2}(\frac{\kappa \Gamma }{2}+\delta
\Delta _{a}-g^{2}N)/(2\kappa \Xi \delta _{0}^{2}|a_{in}|^{2})-1$. It is
obvious that at $\sin {\varphi }=0$, the system demonstrates the perfect
photon absorption, but at $\cos {\varphi }=c$ no perfect photon absorption
is shown, even though the equal output light intensities have been present.
In addition, some parameters including the driving, the detuning and so on
could lead to $|c|>1$, so $\cos {\varphi }=c$ can't hold any more. This case
is illustrated in Fig. 5 (a) where the input intensity $\left\vert
a_{in}\right\vert ^{2}$ is set below the threshold of the bistable regime,
hence $\left\vert a_{1out}\right\vert ^{2}$ and $\left\vert
a_{2out}\right\vert ^{2}$ only show the mono-stability. On the contrary,
when we adjust the parameters to satisfy the condition $|c|\leq 1$, the
output light intensities could show the bistability, which is plotted in
Fig. 5 (b). In this case, the phase can be used to manipulate the
bistability and obtain various complicated bistable patterns in a certain
region.
\begin{figure}[tbp]
\centering
\includegraphics[width=0.8\columnwidth,height=2.5in]{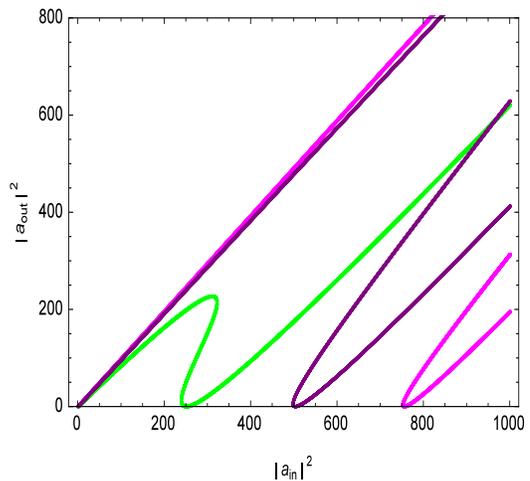}
\caption{(color online). The output intensity $\left\vert a_{out}\right\vert
^{2}$ versus the variable $\left\vert a_{in}\right\vert ^{2}$ with the
different values of $\Delta _{a}/\protect\kappa $. The green, purple,
magenta lines correspond to $\Delta _{a}/\protect\kappa =-5$, $\Delta _{a}/%
\protect\kappa =0$, $\Delta _{a}/\protect\kappa =5$, respectively. The other
parameters are same as Fig. 2. }
\end{figure}
\begin{figure}[tbp]
\centering
\includegraphics[width=0.55\columnwidth,height=2.2in]{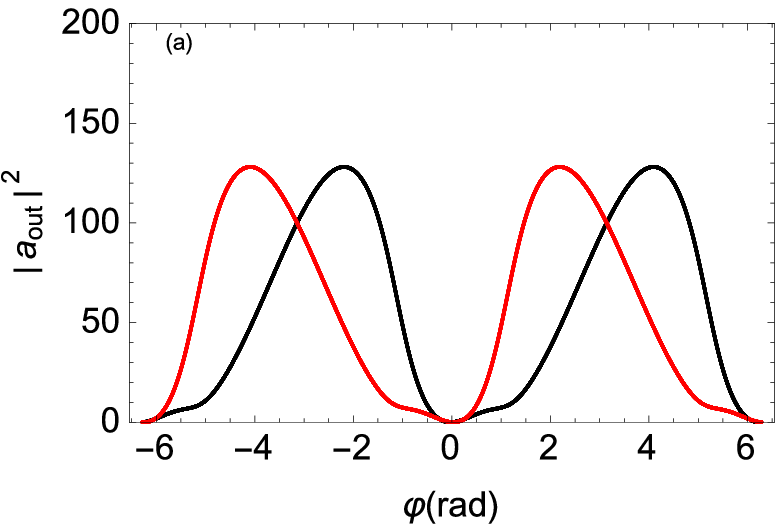}%
\includegraphics[width=0.5\columnwidth,height=2.2in]{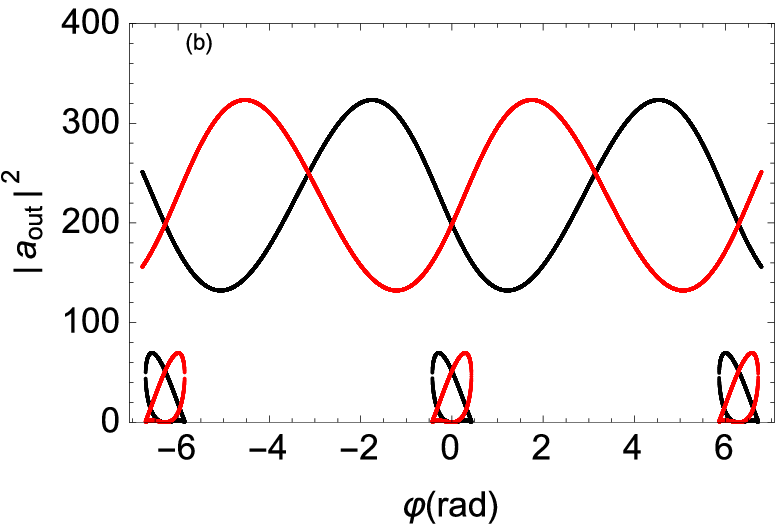}\newline
\caption{(color online). The output intensity $\left\vert a_{out}\right\vert
^{2}$ as a function of the relative $\protect\varphi $. In (a), $\left\vert
a_{in}\right\vert ^{2}=100$, $\Delta _{a}/\protect\kappa =-8$; in (b) $%
\left\vert a_{in}\right\vert ^{2}=250$, $\Delta _{a}/\protect\kappa =-5$.
The red line is for $\left\vert a_{out}^{1}\right\vert ^{2}$ whereas the
black line stands for $\left\vert a_{out}^{2}\right\vert ^{2}$. The other
system parameters are same as Fig. 3. }
\end{figure}

\subsection{Perfect photon absorption beyond the low-excitation limit}

The consequences within the low-excitation limit have been studied in
the previous section. Drowning on the findings above mentioned,
a natural question will be arisen, thus, whether stronger and more robust nonlinearity
can be achieved when the atomic assemble is driven into the nonlinear
excitation regime? we expect that the nonlinearity caused by the atomic nonlinear excitation can be used
to enhance the cavity optical nonlinearity, or has a positive effect on the
perfect photon absorption in the hybrid atom-optomechanical system.

In the following, we will give the details on the coupling between the atomic ensemble and 
the cavity field in the nonlinear regime. Thus, in Eq. (\ref{non}), we will
have to directly employ $\left\langle S^{z}\right\rangle =-\frac{N}{2(1+\frac{2g^{2}\left\vert
a\right\vert ^{2}}{\frac{\Gamma ^{2}}{4}+\delta ^{2}})}$ instead of  $\left\langle S^{z}\right\rangle=-%
\frac{N}{2}$ to calculate the steady-state solution of the intra-cavity
field from the Eq (\ref{lan1})- Eq (\ref{lan2}). At this moment, the $\left\langle S^{z}\right\rangle $ is a
function of the intra-cavity field intensity $\left\vert a\right\vert ^{2}$ which satisfies the
 condition $\left\vert a\right\vert ^{2}>\frac{\Gamma ^{2}}{4g^{2}}$. Therefore, two output optical intensity $\left\vert
a_{1out}\right\vert ^{\prime 2}$ and $\left\vert a_{2out}\right\vert^{^{\prime }2}$ can be given by
\begin{eqnarray}
\frac{\left\vert a_{1out}\right\vert ^{\prime 2}}{\left\vert
a_{in}\right\vert ^{2}} =\left\vert \frac{\kappa (1+e^{i\varphi })}{%
i\Delta _{a}+\kappa +\frac{g^{2}N}{(\frac{\Gamma }{2}+i\delta )(1+\frac{%
2g^{2}\left\vert a\right\vert ^{2}}{\frac{\Gamma ^{2}}{4}+\delta ^{2}})}%
-i\Xi \left\vert a\right\vert ^{2}}-1\right\vert ^{2},  \label{nonlinear} \\
\frac{\left\vert a_{2out}\right\vert ^{\prime 2}}{\left\vert
a_{in}\right\vert ^{2}} =\left\vert \frac{\kappa (1+e^{i\varphi })}{%
i\Delta _{a}+\kappa +\frac{g^{2}N}{(\frac{\Gamma }{2}+i\delta )(1+\frac{%
2g^{2}\left\vert a\right\vert ^{2}}{\frac{\Gamma ^{2}}{4}+\delta ^{2}})}%
-i\Xi \left\vert a\right\vert ^{2}}-e^{i\varphi }\right\vert ^{2}.  \label{nonlinear2}
\end{eqnarray}

In order to show the perfect photon absorption, we would like to first list the corresponding conditions similar to the case in low-excitation limit. With $\phi=0$, the conditions parallel with Eqs. (\ref{perfect11}) and (\ref{perfect12}) can be
straightforwardly written as
\begin{eqnarray}
\frac{\kappa \Gamma }{2}+\Delta _{a}\delta &=&\frac{g^{2}N}{1+\frac{%
2g^{2}\left\vert a\right\vert ^{2}}{\frac{\Gamma ^{2}}{4}+\delta ^{2}}}%
+2\delta \Xi \left\vert a\right\vert ^{2},  \label{perfect21} \\
\frac{\Delta _{a}\Gamma }{2}-\kappa \delta &=&\frac{\Gamma \Xi }{2}%
\left\vert a\right\vert ^{2}.  \label{perfect22}
\end{eqnarray}
Compared with  the perfect absorption conditions of
the CQED system in Ref. \cite{Agarwal-arxiv}, one can find that both the nonlinear
cavity-mechanical coupling and the atom-cavity coupling have a certain influence  on the perfect
photon absorption in the
nonlinear coupling regime.

In addition, from the Eq. (\ref{non}), one can find
that the intra-field intensity $\left\vert a\right\vert ^{2}$ satisfies a
quintic equation. Let $\left\vert a_{in}^{1}\right\vert
^{2}=\left\vert a_{in}^{2}\right\vert ^{2}=\left\vert a_{in}\right\vert ^{2}$
$(\varphi =0)$ and  $x:=$ $\left\vert a\right\vert ^{2}$, the quintic equation
reads
\begin{equation}
A_{5}x^{5}+A_{4}x^{4}+A_{3}x^{3}+A_{2}x^{2}+A_{1}x+A_{0}=0,  \label{quintic}
\end{equation}%
where
\begin{equation}
A_{0}:=-\left\vert \tilde{\Gamma}\right\vert ^{2}\left\vert
a_{in}\right\vert ^{2},
\end{equation}%
\begin{equation}
A_{1}:=(\tilde{\kappa}\tilde{\Gamma}+g^{2}N)(\tilde{\kappa}^{\ast }\tilde{%
\Gamma}^{\ast }+g^{2}N)-2\left\vert \tilde{\Gamma}\right\vert ^{2}\tilde{%
\delta}\left\vert a_{in}\right\vert ^{2},
\end{equation}%
\begin{eqnarray}
A_{2} &:&=\tilde{\Gamma}\tilde{\delta}(\tilde{\kappa}-i\Xi )(\tilde{\kappa}%
^{\ast }\tilde{\Gamma}^{\ast }+g^{2}N)+  \notag \\
&&\tilde{\Gamma}^{\ast }\tilde{\delta}(\tilde{\kappa}^{\ast }-\tilde{\omega}%
_{m}^{\ast })(\tilde{\kappa}\tilde{\Gamma}+g^{2}N)-\left\vert \tilde{\Gamma}%
\right\vert ^{2}\tilde{\delta}^{2}\left\vert a_{in}\right\vert ^{2},
\end{eqnarray}%
\begin{eqnarray}
A_{3} &:&=\left\vert \tilde{\Gamma}\tilde{\delta}(\tilde{\kappa}-i\Xi
)\right\vert ^{2}-i\Xi \tilde{\Gamma}\tilde{\delta}(\tilde{\kappa}^{\ast }%
\tilde{\Gamma}^{\ast }+g^{2}N)  \notag \\
&&+i\Xi \tilde{\Gamma}^{\ast }\tilde{\delta}(\tilde{\kappa}\tilde{\Gamma}%
+g^{2}N),
\end{eqnarray}%
\begin{equation*}
A_{4}:=i\Xi \tilde{\Gamma}^{\ast }\tilde{\delta}^{\ast }(-i\Xi -\tilde{\kappa%
}^{\ast })-i\Xi \tilde{\Gamma}\tilde{\delta}(i\Xi -\tilde{\kappa}),
\end{equation*}%
\begin{equation}
A_{5}:=\tilde{\delta}^{2}\left\vert i\Xi \tilde{\Gamma}\right\vert ^{2},
\end{equation}%
\begin{equation}
\tilde{\Gamma}:=i\delta +\frac{\Gamma }{2},
\end{equation}%
\begin{equation}
\tilde{\kappa}:=\kappa +i\Delta _{a},
\end{equation}%
\begin{equation}
\tilde{\delta}:=\frac{2g^{2}}{\frac{\Gamma ^{2}}{4}+\delta ^{2}}.
\label{canshu}
\end{equation}
This quintic equation can not be analytically solved, so we only solve it
numerically based on the Abel--Ruffini theorem \cite{jjj,zhang}. The numerical results are plotted in Fig. 6 which reveals a remarkable
behavior of the intra-cavity intensity and the output field intensity.
\begin{figure}[tbp]
\centering
\includegraphics[width=1\columnwidth,height=2in]{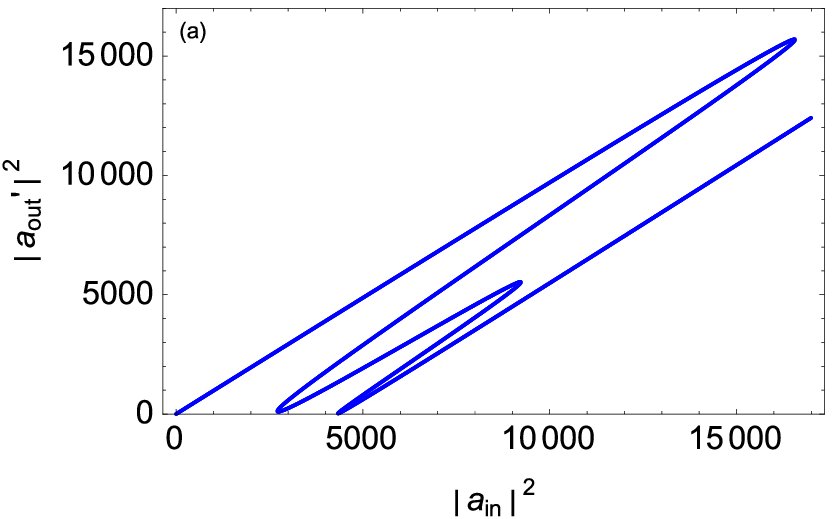} %
\includegraphics[width=1.06\columnwidth,height=2in]{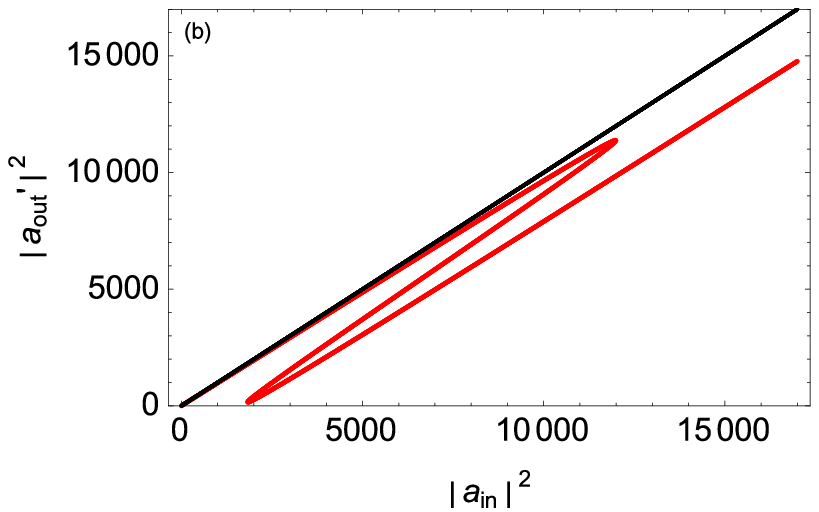}
\caption{(color online).Output light intensity $\left\vert
a_{out}\right\vert ^{2}$ as a function of input light intensity $\left\vert
a_{in}\right\vert ^{2}$ beyond the low-excitation limit. For this case, the
two input lights have the same intensity $\left\vert a_{1in}\right\vert ^{2}$
=$\left\vert a_{2in}\right\vert ^{2}$=$\left\vert a_{in}\right\vert ^{2}$.
In (a), $\Xi/\protect\kappa=0.0065$, in (b), the black line is for $g/%
\protect\kappa=0$, the red curve stands for  $\Xi/\protect\kappa=0$ (absent of the optomechanical interaction). The other systems parameters are as follows: $\Delta _{a}/\protect\kappa =60$, $\protect\delta %
/\protect\kappa=3,$ $g^{2}N/\protect\kappa ^{2}=240$.}
\end{figure}
In Fig. 6 (a), we
show the optical multistability and the perfect photon absorption as well as the relation between them. 
 We observe that with the increasing of the input intensity, the output intensity demonstrates the monostability, bistability and multistability, respectively. One can notice that the onset of multistability requires a stronger input field intensity comparing with
the linear excitation regime. 
This result suggests that the nonlinear coupling of atomic excitation could be
employed to enhance the optical nonlinearity. It is especially interesting that the perfect photon absorption occurs at two particular input intensities: one is $\left\vert a_{in}\right\vert ^{2}=\frac{2g^{2}N\Gamma -\kappa
\Gamma ^{2}-4\kappa \delta ^{2}}{8g^{2}}$ and the other is $\left\vert
a_{in}\right\vert ^{2}=\frac{\kappa \Gamma \Delta _{a}-2\kappa ^{2}\delta }{%
\Gamma \Xi }$. This indicates that one can tune the input field intensity under
the perfect absorption conditions (Eq.(\ref{perfect21}),Eq.(\ref{perfect22})) to
achieve the two perfect photon absorption points. Thus, the perfect photon
absorption can be nonlinearly controlled by the input field intensity.
Furthermore, one can find that the two perfect photon absorption points
correspond to the optical bistability and multistability, respectively.  This
means that the two nonlinearities induced by the atom-cavity coupling and the cavity-mechanical coupling can be overlap-added in a certain
parameters regime. In other words, the nonlinear optomechanical
interaction not only enhances the optical nonlinearity but also adds an additional perfect
photon absorption point.

What's more, we would like to emphasize  the main difference between the current
scheme and the previous work \cite{Agarwal-arxiv}. The hybrid atom-optomechanical system provides
a further understanding for the corresponding relation between the optical
bistability/multitability and the perfect photon absorption. This can be  shown
by an example in Fig. 6 (b). The red line corresponds to $g_{0}/\Gamma =0$
(without the cavity-optomechanical coupling), whereas the black line corresponds to
the standard optomechanical system without the atomic interaction.
In Ref. \cite{Agarwal-arxiv}, we find that the
nonlinearity arises from the atomic nonlinear excitation regime which plays the
dominate role in the optical bistability. In our case, when the atom-optomechanical
coupling is absent, the hybrid optomechanical system is reduced to
the CQED system, which is just consist with  Ref. \cite{Agarwal-arxiv}. Once we
consider the optomechanical coupling, one can find the optical nonlinearity has been enhanced and the perfect photon
absorption occurs at two special   input intensities $\left\vert
a_{in}\right\vert ^{2}$ as shown by
the blue line in Fig. 6 (a). Similar to the case of the low-excitation limit, one can find that, if $g=0$ (in
absence of the atom ensemble), both the perfect photon absorption and the
optical bistability of the output intensity  disappear. This can be
explained as follows. The output field intensity is given by $\left\vert
a_{out}\right\vert ^{\prime 2}=\left\vert \frac{\kappa -i\Delta _{a}+i\Xi
\left\vert a\right\vert ^{2}}{i\Delta _{a}+\kappa -i\Xi \left\vert
a\right\vert ^{2}}\right\vert ^{2}\left\vert a_{in}\right\vert
^{2}=\left\vert a_{in}\right\vert ^{2}$ (here we set $\varphi =0$) which implies
that the
output field intensity still behaves with the monostable properties, even though the intra-cavity field is driven in the bistable regime. This is analogous
to the optomechanically induced transparency (OMIT) \cite{omit,science330}. Equivalently, one can draw 
the same conclusion directly from the violation of the perfect absorption conditions given in Eq. (\ref{perfect21})
 and Eq. (\ref{perfect22}).

Before the end, we briefly discuss the experimental feasibility of our scheme.
 Our proposal mainly depends on the parameters $g\sqrt{%
N},\kappa ,\Gamma $ and $g_{0}$. In experiment, $g\sqrt{N}/\Gamma $=20 can
be realized by the ultra cold atomic ensemble coupled with the cavity
with the atomic half-linewidth $\Gamma $= $2\pi \times 3M$ Hz \cite%
{Bren,omshiyan,shiyan1}. In addition, the optomechanical parameters we used
can be realized in various optomechanical systems, for example, $\omega
_{m}/2\pi =4.2\times 10^{4}$ Hz, $g_{0}/2\pi =6\times 10^{5}$ Hz, $\kappa
/2\pi =6.6\times 10^{5}$ Hz have been reported in Ref. \cite{modern}, so the values of $%
\Xi $ used in the main text can be easily achieved. To sum up, one can 
find that all the conditions required in this paper are realizable within the current experimental
technology.


\section{Conclusions and discussion}

In summary, we have studied the optical  response properties and the suppression of the output fields in the hybrid atom-optomechanical system with  two identical
input laser fields. It is shown that  the perfect photon absorption is present in  both the  linear and  nonlinear atomic excitation
regimes.  We find that in such a hybrid optomechanical system, the coupling with atomic ensemble is the sufficient and necessary condition for the perfect photon absorption, while the optomechanical coupling, enhancing the nonlinearity, forms the necessary condition for the optical bistability/multistability and  especially induces an additional perfect absorption point in the nonlinear atomic excitation regime. As a result, we also find that the perfect photon absorption corresponds to the optical bistability/multistability. Furthermore, one can find that the interference of the two input fields can modify the photon absorption and
optical bistability (multistability), so the bistability (multistability) and perfect
absorption can be controled by changing the relative phase between them, which
provides a possible design of an optical switch.
Finally, one should note
that all the parameters employed in the numerical procedure are taken from the practical experiments, which ensures
the practical feasibility  in a variety of COM systems.

\begin{acknowledgments}
This work was supported by the National Natural Science Foundation of China,
under Grant No.11375036 and 11175033, the Xinghai Scholar Cultivation Plan
and the Fundamental Research Funds for the Central Universities under Grants
No. DUT15LK35 and No. DUT15TD47.
\end{acknowledgments}

\end{document}